\documentclass[conference]{IEEEtran}
\IEEEoverridecommandlockouts
\usepackage{amsmath,amssymb,amsfonts}
\usepackage{algorithmic}
\usepackage{graphicx}
\usepackage{textcomp}
\usepackage{xcolor}
\usepackage{hyperref}

\def\BibTeX{{\rm B\kern-.05em{\sc i\kern-.025em b}\kern-.08em
    T\kern-.1667em\lower.7ex\hbox{E}\kern-.125emX}}
\usepackage[style=ieee]{biblatex} 
\begin{document}

\title{Analysis of reward mechanism for quizmarket\\
{\footnotesize \textsuperscript{*}}
\thanks{}
}

\author{\IEEEauthorblockN{Noorul Ali \footnote}
\IEEEauthorblockA{\textit{Computer Science and Engineering} \\
{IIIT Vadodara}\\
Gujarat, India \\
201851078@iiitvadodara.ac.in}
}

\maketitle

\begin{abstract}
A reward algorithm is needed for games which rewards risk, i.e. early play, and extends the longevity of a reward pool. This would allow a higher number of players and greater engagement. I created a reward mechanism that rewards risk, lasts longer, and is more profitable than existing mechanisms. I also implemented an algorithm within the mechanism to self-correct in outlier performance. This reward mechanism was used in TURBLAZE, a mobile game designed for high school students. The game has quizzes. Gamers pay a fixed fee to participate in a quiz and win a reward if their score is above a certain threshold. 
\end{abstract}
\footnotetext[1] {This work was carried out as part of research internship under the mentorship of Pratik Shah, \textit{Associate Professor}, IIIT Vadodara (May-July 2021)}

\section{Introduction}
This algorithm was designed specifically with a mobile game in mind. TURBLAZE is a mobile game I designed to increase student participation in curriculum by making a mobile video game for high school education.

Quizmarket is a marketplace in this game for quizzes, where details such as reward, participation fee, difficulty level, reward history etc. are shown to the gamer.

Each quiz has a fixed prize pool which is to be distributed appropriately to winners. The reward mechanism we have made is such that the number of winners is maximized, while maintaining hierarchy in winners. Hierarchy is such that being early is rewarded.

The quiz also has to ensure that all expenses incurred, including prize pool and hosting expenses, are less than the money made from the fee. The mechanism takes into account all kinds of expenses and delivers a way to host a profitable quiz while maximizing the number of participants. It is also fair to all participants.

\section{Literature Survey}

\subsection{Existing reward mechanisms for games}

Other games that have reward mechanisms use randomness to create fairness in gambling. Examples of these include fantasy sports games \cite{b6}, online rummy \cite{b7}, and betting sites \cite{b8}.
Pre-set probabilities ensure that the game makes money. This simplifies the game significantly, as randomness within range introduces luck, which works in favor for these games. These games are not just about skill, and their algorithms reflect that. But we need knowledge to be source of rewards, and believe luck should not deter a knowledgeable person. 

\subsection{Algorithms in casinos}

A normal odd-even betting game in which getting an even number is a win and an odd number is a loss has a probability of 50-50. Roulette adds two numbers into the losing pile and shifts the probability slightly to ensure the casino always makes money. Almost all games are ultimately designed to make the casino a profit, when the game is played at scale. 

Casinos build their foundations on luck. It is always the element of randomness and luck that makes it money. From small village gambles (\textit{satta}) to large casinos in Vegas, games are always rigged to make the house money. We are inspired by these algorithms but the randomness of a dice roll too often overpowers skill.

\subsection{Blockchain gambling pools}

Betting pools have emerged on blockchain technology \cite{b4,b5,b9}. These bets are made with many people in a pool, and upon completion of bet condition, a winner is chosen at random. There is an inherent lack of instant personal feedback, due to randomized distribution of prize. 

Bitcoin reward for each mined block halves roughly every 4 years, when a particular number of blocks is mined. This ensures that the overall supply of bitcoin tokens remains fixed at 21,000,000. Inspiration is taken from this infinite geometric progression, because prize pool for every quiz is also fixed.  

\section{Problem Formulation}
Each game is a 10 minute quiz. Score above threshold, win reward. Else, get online resources to learn what you got wrong.

Each quiz has a fixed prize pool (pp). This is initial capital. The reward (r) for winners comes from this pool. This pool is distributed in such a way that numbers of winners are  maximized and early winners score more. Number of winners is directly proportional to number of registrations, and hence directly proportional to profitability of quiz.  

\subsection{Constants}
Registration fee (f) is constant. This is the fee that a user gambles on and pays to participate in quiz. This used to make quiz profitable, all expenses are less than the total fee collected. 

Cost/user (c): this is the cost incurred per game per user. Registration, cloud expenses, etc. are included in this. It is formulated as a fixed percentage of prize pool. 

Hosting cost (h): this is company expenses for hosting quiz, formulating questions etc. Also formulated as a fixed percentage of prize pool 

Initial prize pool (ipp): This is the total prize pool allotted to quiz before it starts. It is fixed and rewards are subtracted from it. Its use needs to be maximized.  

\subsection{Variables}

Threshold (t): every quiz has a winning score. If a student scores above this score, they win the reward. This is variable so that the difficulty of quiz varies as per the performance of gamers. 

Prize pool (pp): this is the prize pool left after subtracting rewards.

Cost-percentage (cp): this is the percentage of registration fee kept by game. This is used to meet expenses and make quiz profitable. It is variable according to the party hosting the quiz. A higher percentage allows quicker cost recovery but limits the number of winners, hence longevity of quiz.  

Reward(r): this is the reward given to a user upon scoring higher than the threshold. It is variable such that risk taking of early winners is rewarded. This is deducted from prize pool (pp). 

\subsection{Core Ideas}
\begin{itemize}
\item  Threshold depends on success rate and existing prize pool, so that more difficult quizzes are incentivized
\item  Prize pool varies as registrations increase, since a percentage of fee goes to pool. This means cp less than 1.
\item  Reward must be based on success rate and current prize pool
\item  Quiz is viable only till reward greater than fee
\end{itemize}

\section{Proposed Solution}

\subsection{Case 1 algorithm: Limited case with simple GP}

Prize pool (pp) is assumed to be constant. Cp=100\%, the entire registration fee (f) of every user is taken by quiz as expense. Upon winning, 
\begin{equation}
pp = pp – r 
\end{equation}

Follows GP for reward, sum of rewards = prize pool
\begin{equation}
Rewards: a,ar,ar^2 … ar^n
\end{equation}
\begin{equation}
Reward(nth) = ar^n
\end{equation}
\begin{equation}
n \rightarrow infinity
\end{equation}
For r = 0.9695, n is maximum while r (reward) greater than f (fee)

\begin{equation}
n = 37
\end{equation}

\subsection{Limitations of case 1 algorithm}
Constant prize pool limits number of winners. This is problematic since longevity and consequent profitability of quiz is affected. 

\subsection{Case 2 algorithm: optimized GP}

Prize pool (pp) is assumed to be variable. Cp = 75\%. Thus 25\% of f (registration fee) is added to prize pool to increase longevity. 75\% of f (fee) is taken as expense. For every registration, 
\begin{equation}
pp = pp + 0.25f 
\end{equation}
Upon winning, 
\begin{equation}
pp = pp – r 
\end{equation}
In geometric progression of reward
\begin{equation}
pp=a(1-x^n)/1-x\label{eq}
\end{equation}
\begin{equation}
Constant = pp
\end{equation}
\begin{equation}
Variables = x , n
\end{equation}

Result:
\begin{equation}
n = 46
\end{equation}

Changing cp increases the maximum number of winners by 24\%, allowing a comfortable profitability ratio for standard 20\% win-ratio.

\subsection{Limitations of case 2 algorithm}
Algorithm does not modify itself to account for everyone winning, can lead to losses

\subsection{Case 3 algorithm: self optimizing GP}

This builds on case 2 algorithm. Threshold (t) corrects itself based on success rate. A quiz with too many winners has a higher threshold, becoming more difficult to win, and effectively decreasing the number of winners to maintain the pre-set win-ratio.

Initial threshold is t\_0, which is the standard pre-set threshold that is to be maintained. t is the updated threshold, which updates itself after every 5 quiz completions.  
For a 70\% success ratio, 
\begin{equation}
t\_0 = 0.7
\end{equation}
Correcting factor (cf):
\begin{equation}
cf = (1/5 - s)sf
\end{equation}
1/5 = 0.2 = average success rate needed

s = success rate, changes after every 5 quiz completions

sf = perturbation rate (scaling factor)

For sf = 10,

\begin{equation}
t = 10(1/5-s) + 70
\end{equation}
Threshold lies between 60\% and 95\%

Case 3 algorithm varies difficulty within reasonable limits to account for all types of users, ensuring game keeps making money

\section{Interpretations}
\subsection{Basic assumptions}
If ipp is set to \$100, with f = \$1 for every user,

\subsection{Case 1 results}
cp = 100\%, all of f (fee) is taken as expense. The maximum number of winners achieved in this case is 37. For a 20\% success ratio, this allows for ~185 registrations.
\subsection{Case 2 results}
cp = 75\%, \$0.75 of f (fee) is taken as expense. \$0.25 is added to pp on each registration. The maximum number of winners achieved in this case is 46. This is a 24\% increase. For a 20\% success ratio, this allows for ~250 registrations. These winners must be interspersed to allow 250 registrations. In case they occur together, the prize pool will get exhausted quickly.

\subsection{Case 3 results}
In this, threshold optimizes itself to ensure scalability. It maintains 46 winners no matter how they occur in registrations. Winners need not be interspersed. Ensures profitability at scale (20 gamers to 20000 gamers).

\section{Experimental discussions}
Scalability of this algorithm has been tested. It works to generate rewards in appropriate geometric progressions, from 20 users to 20000. Various tools to generate backend infrastructure were discussed upon. 

\section{Further developments}
This reward mechanism can be used in any kind of task/game where there is a limited resources have to be effectively distributed on the basis of a certain parameter like skill. Using this, the architecture for mobile game TURBLAZE was created. 

\section{Conclusions}
Inspired from casinos and Bitcoin \cite{b1,b4}, rewards are a geometric progression that ensures maximum number of winners for a particular pool. This allows for maximum utilization of pool, while reward is profitable compared to registration fee. The reward also fluctuates on the basis of success rate of quiz (which determines quiz difficulty in real-time). Finally, in the spirit of casinos, the game is always rigged to make TURBLAZE money. No matter the number of winners, the quiz difficulty and innate (by design) structure that fluctuates reward on the basis of success rate, ensures profitability at every scale, from 20 players to 20000. 

\section{Thanks}
I am extremely grateful to Pratik Shah sir for listening patiently as I rambled on and on about far off goals, while I delayed my submissions. Thank you for the clarity and giving me insights. They helped me a lot. The discussions with you were extremely fruitful and helped me achieve important goals I was unaware of. I used to only write code, but this internship taught me what it means to make a project, and I am very thankful for that. Thank you once again sir.

\printbibliography

\end{document}